\documentclass{cernyrep}
\usepackage{texnames}
\usepackage[T1]{fontenc}
\usepackage{eurosym}
\usepackage[bookmarks, colorlinks=true, linktoc=page, pdftex, linkcolor=black, citecolor=black, urlcolor=blue]{hyperref}
\usepackage{fancyhdr}
\fancyhfoffset{4 mm}
\fancypagestyle{ARTTITLE}{%
\fancyhf{} 
\lhead{\small{Proceedings of the 2018 CERN--Accelerator--School course on \it{Beam Instrumentation}, Tuusula, (Finland)}}
\lfoot{Available online at \url{https://cas.web.cern.ch/previous-schools}}
\rfoot{\thepage\hspace*{3mm}}

}

\begin{document}
\title{Diagnostic Needs for Wakefield Accelerator Experiments}

\author {A. Cianchi}

\institute{Universit\`a degli Studi di Roma "Tor Vergata", Roma, Italy
}

\begin{abstract}
Wakefield accelerators are under development in many laboratories worldwide. They bring the promise of a high accelerating gradient, orders of magnitude higher than current machines. The reduction in the overall length of the accelerators will pave the way to a wider use of such machines, for industrial, medical, research, and educational purposes. At the same time, all the equipment must be reduced as well, to keep the dimensions of the machine as small as possible.
The two main challenges of the diagnostics for plasma accelerated electron beams are the ability to measure the 6D phase space properties with single shot techniques and the compactness to meet the requirements of a `table-top' facility.
\end{abstract}


\maketitle 

\thispagestyle{ARTTITLE}

\section{Introduction}

In 1946, ENIAC, one of the first computers in  history, occupied an area of 180 square metres with a weight of 30 tons.
Forty-nine years later the same calculation power was achieved by a chip of 7.44 X 5.29 mm$^{2}$.
It was 1995 and the chip, working at 20 MHz, is now in the Computer History Museum in San Jose, because now, even our smartphone has a processor working at a much faster clock.
Try to imagine being in 1946 and  predicting the day an enormous computer would be reduced to the size of a finger tip.
It is a high mountain to climb, and not one climbed in a day but in half a century.
We can consider this case an example of the ongoing transformation in the field of particle accelerators.

The reason why nowadays the high energy, e.g., GeV scale, accelerators are km long is due to the low accelerating gradient, the same order of magnitude as that of about 50 years ago.
To achieve higher and higher energies the only way has been the increase of the overall machine length.

Why is the accelerating gradient so limited?
The charges are accelerated in metallic structures, where a high electric field is established.
Above a certain level, a mechanism, called breakdown, appears \cite{bib:grudiev}.
Electrons are extracted from the metallic walls and driven on an arc discharge to another point in the structure, dissipating at a single point a large quantity of energy and likely damaging the metallic structure itself.
Practical values of maximum achievable field are today of the order of 100--150 MV/m, established only in prototyping accelerators in the X-band \cite{bib:dalforno}.

Plasma acceleration foresees fields of the order of tens or hundreds of GV/m, allowing us to reach high energies in very short spaces.
While this lecture is not devoted to a comprehensive treatment of the physics of plasma accelerators, for which there is also a CERN Accelerator School proceedings already published \cite{bib:cas}, to improve the comprehension of the readers and to understand better which are the problems that we face in beam diagnostics, some main concepts and schemes are introduced in the following chapter.

\subsection{Wakefield acceleration}

A simple definition of a plasma is a quasi-neutral gas of charged particles showing collective behaviour.
Quasi-neutral means that the number densities of electrons and ions are locally balanced.
In a collective behaviour the long-range  Coulomb potential usually dominates over microscopic fluctuations.
There are different kinds of plasmas, some of them with low temperature and electron density, like interstellar plasmas, some others with very high temperature and electron density, like plasmas for nuclear fusion \cite{bib:plasma}.
Plasmas for particle accelerators are in between these two limits, showing densities of around 10$^{15}$-10$^{19}$ cm$^{-3}$.

How does plasma acceleration work?
An electric field is needed to accelerate charges.
If we consider a tightly focused  high intense laser, about 10$^{18}$W/cm$^{2}$ in a few $\mu$m spot, over a supersonic gas jet, the pulse strips off the electrons in the gas \cite{bib:self}, producing a plasma.
The ponderomotive force of the laser bullet is high enough so that the much lighter electrons are blown outward in all directions, leaving behind the more massive ions.
When they reach the laser pulse propagation axis, they overshoot it and end up travelling outward again, producing a wave-like oscillation.
The electrons actually form a bubble-like structure, in front of which there is the laser pulse that creates the plasma.
This scheme is called laser wakefield acceleration (LWFA).

With the number of charges very large, the electric field that appears, due to the unbalancing of the positive and negative charges, can be so large that some electrons can be self-injected in the rear part of the bubble.
There they experience a strong accelerating field.
This scheme is called self-injection because the electrons are injected by the surrounding plasma, without the need for any external source.
With this system, beams with an energy of up to 8 GeV have been produced in just 20 cm \cite{bib:8gev}.
However, due to the lack of any real control in the injection process, the quality of the beam is not suitable for high level applications, like for instance free-electron lasers (FEL) and colliders.

\begin{figure}[h]
  \centering
  \includegraphics[width=60mm]{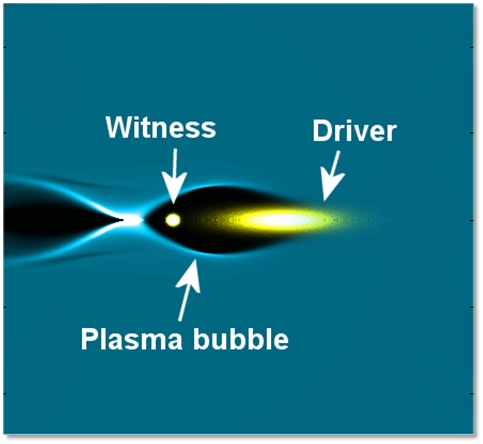}
  \caption{A high brightness electron beam (or high intensity laser) drives a plasma wave. A witness beam injected on the back of the bubble is then accelerated.}
  \label{fig1}
  \end{figure}

A better control of the accelerated beam can be obtained with a scheme which uses particle beams instead of lasers in order to excite plasma waves \cite{bib:chen}.
This scheme is called beam-driven plasma wakefield acceleration (PWFA) and it is briefly illustrated in Fig.\ref{fig1}.

A high brightness electron beam (driver) is injected inside the plasma.
The space charge force repeals electrons creating a bubble structure of unbalanced charges.
A witness bunch is injected in the rear part of the bubble and is accelerated by the intense electric field.
With the witness produced by a conventional accelerator, its properties can be properly tailored to obtain the best emittance and energy spread at the end of the plasma acceleration.
However, the beam quality depends heavily on the ability to properly tailor the driver and witness phase spaces at the same time.

To overcome this problem it is possible to replace the driver beam with a laser beam \cite{bib:extinj}.
In this case, the accelerator has to produce only the witness beam.
Because the driver is the laser, the accelerator can be tuned to have the best witness beam parameters.
The most relevant difficulty of this scheme is the synchronization between laser and electron beam.
With a laser pulse of the order of few tens of fs, the jitter between the two must be much lower than this value.

Despite  the scheme used in plasma acceleration, from the diagnostics point of view we have to measure the beam leaving the plasma stage (output beam, in all  cases) and also the beam entering in the plasma stage (input beam, all but self-injection).

\section{Diagnostics general remarks}

Several parameters must be measured with remarkable resolution and possibly single shot capability.
The requirement of a single shot  comes mainly in the commissioning phase of these kinds of accelerators, where a mismatch of the beam, a problem with laser guiding, a jitter between a laser and the beam, a pointing instability in the produced beam, or an energy jitter, can produce very different beams shot by shot.
The optimization could be very hard without a single shot capability.
But also the resolution is of paramount importance.
Longitudinal diagnostics must measure bunch lengths of the order of a few fs.
Transverse emittance can be of the order of 1 mm-mrad or a fraction of this value.
Bunch charge can be as low as a few pC.
Having at the same time single shot measurement and high resolution is even more complicated.

Even conventional diagnostics are hard to  implement in such machines.
The compactness of the overall machine is an indispensable quality.
It is not possible to have structures of a few centimetres, able to accelerate the beam up to GeV level, and later use metres of space to measure the beam properties.

Another particular feature of plasma accelerators is the difficulty in finding the proper space for the diagnostics.
There are several considerations preventing the placement of diagnostics just after the accelerating module.
First of all, the driver must be removed, be it a laser or electron beam.
The laser can destroy or severely damage all the devices downstream of the plasma module, while the electron beam driver can interfere in the beam witness properties.
A chicane, a laser mirror, which is a dedicated optics scheme must be implemented before placing any device to measure the beam parameters.
This is valid both in commissioning and in the normal operation case.

It is not only a geometrical problem.
It has been noticed that the normalized root mean square (rms) emittance can grow even in a drift in the presence of energy spread \cite{bib:flot,bib:mauro}.
We recall that the total transverse normalized emittance squared is:
\begin{eqnarray}\label{e1}
\varepsilon _n ^2  = \left\langle \gamma  \right\rangle ^2 \sigma _\varepsilon ^2 \left\langle {x^2 } \right\rangle \left\langle {x'^2 } \right\rangle  + \left\langle {\beta \gamma } \right\rangle ^2 \left( {\left\langle {x^2 } \right\rangle \left\langle {x'^2 } \right\rangle  - \left\langle {xx'} \right\rangle ^2 } \right)\, ,
\end{eqnarray}
where $\gamma$ is the usual relativistic factor, $\beta$ is the ratio of the speed of the particle to the speed of light, and $\sigma _\varepsilon$ is the percentage energy spread.

Due to the presence of a non-negligible energy spread, and with the divergence term usually of the order of mrad, the first term will be the leading one after some drift.
At this point the normalized emittance has grown significantly, spoiling the beam properties.
Recently, the concept of chromatic length  \cite{bib:conti} has  been introduced, defined as the distance where the emittance grows by a factor $\sqrt 2$ as

\begin{eqnarray}\label{e2}
L_C  = \frac{{\sigma _x }}{{\sigma '_x \sigma _E }}\, ,
\end{eqnarray}

where $\sigma _x$ is the rms beam size, $\sigma '_x$ is the rms beam divergence, and $\sigma _E$ is the relative rms energy spread at plasma extraction.
In a conventional accelerator $L_{C}$ is usually longer than the whole machine, while in plasma accelerators, depending on the value of the energy spread, it could be  between a few centimetres and few metres.
To overcome this problem the only solution is a fast capture of this beam and a mitigation of the energy spread, even at the cost of some charge reduction.
This really prevents the possibility of installing diagnostics just after the plasma structure, if the energy spread is not reduced to the level of conventional accelerators, of the order of about 0.1$\%$

\section{Input beam}

\begin{figure}[h]
  \centering
  \includegraphics[width=60mm]{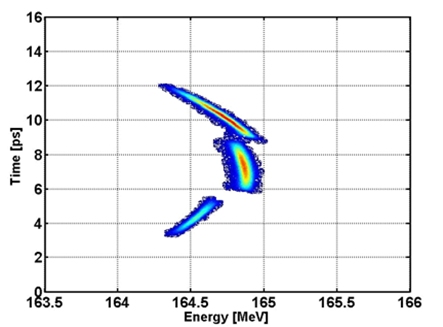}
  \caption{Longitudinal phase space of a train of bunches, credits:\cite{bib:cianchi} }
  \label{fig2}
  \end{figure}

Let us now consider the case of PWFA, with a driver beam to excite the plasma wake.
We have a train of at least two bunches, one driver and one witness.
Also, it is possible to produce multiple driver beams, followed by a single witness bunch, to increase the accelerating gradient.
The energy transferred to the plasma is in this case increased by the number of  drivers.
Usual values of the time distance between bunches is of the order of ps or a fraction of this.
A real 6D diagnostic must separate in time the two bunches and must be able to measure the parameter of every single bunch of the train \cite{bib:cianchi}.
For longitudinal measurement a transverse deflecting structure (TDS) \cite{bib:behrens}, often called a radio-frequency deflector (RFD) \cite{bib:emma} can separate the bunches and  can allow the measurements of the bunch length and bunch distance.
The use of a dipole allows us to have in the same single shot the dispersion in time and in energy, resulting in a picture of the longitudinal phase space, as  shown in Fig. \ref{fig2}.

So the use of a TDS allows single shot measurement of all of longitudinal parameters.
However, the resolution of this measurement is fundamental, because the bunch length can be of the order of a fraction of ps, down to a few fs.
The resolution of a TDS is inversely proportional to the RF wavelength and beam energy and directly to the deflecting voltage.
To achieve fs resolution at GeV energy, an X-band TDS is required.
This device is not yet fully developped and a new design and implementation are under way.
While this device can reach such a resolution, particular attention must be paid to  its design.
The reduced iris aperture and the possibility that the beam goes off centre inside the device, due to the transverse field, must be considered with beam dynamics simulations.
Only one X-band RFD is operating so far at SLAC \cite{bib:xband}.
Particular attention must be paid also to RF time jitter and the structure temperature stability, tailoring these requirements to the beam energy and the available beam line.
As a rule of thumb, for a GeV level beam,  temperature stability of the order of tens of mK and time stability of the order of tens of femtoseconds are required.

\begin{figure}[h]
  \centering
  \includegraphics[width=60mm]{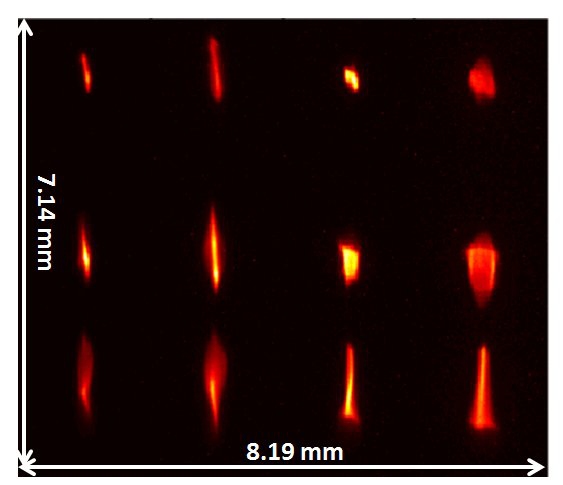}
  \caption{A train of two drivers and one witness separated by a TDS. The picture is a collection of four different measurements, each of them with a different value of the current in a quadrupole placed between the TDS and the screen.}
  \label{fig3}
  \end{figure}

For the transverse phase space a conventional technique, like for instance quadrupole scan \cite{bib:lohl}, cannot give the value of the emittance of the single bunch in the train.
But the use of a TDS can solve the problem.
 Figure \ref{fig3}  shows four different measurements of a train of three bunches, with four different values of a quadrupole in the drift between the TDS and a screen.

A quadrupole scan can be done for the orthogonal plane with respect to the deflecting voltage.
So if we use a TDS with vertical deflection, we can measure the emittance with the quadrupole scan in the horizontal plane.
The TDS acts like a drift in the plane orthogonal to the deflection.
Extracting the transverse parameter from each single bunch, the emittance of every bunch in the train can be measured \cite{bib:cianchi2}.
Today, a new TDS with the possibility of  rotation of the deflecting mode \cite{bib:grudiev2} is under test.
With such a device it will be possible to streak the beam both in the vertical and horizontal directions, measuring beam emittance in both planes.

\section{Output beam}

We have seen that the measurement of the longitudinal and transverse properties of a train of bunches is possible, even with a THz repetition rate, with existing techniques.
However, it is much harder to apply conventional diagnostics to what we called the output beams,  plasma accelerated beams.

\subsection{Transverse measurement}

\subsubsection{Pepper pot}

A popular approach, until few years ago, was the use of a pepper pot, see for instance Refs. \cite{bib:fritzler},\cite{bib:sears},\cite{bib:brunetti}.
This technique is very well described in detail in Ref. \cite{bib:cianchi2}.
Briefly, the beam is stopped or heavily scattered by a mask of a material of high atomic number, usually tungsten, while some parts of the beam, called beamlets, pass through small holes or apertures.
The beamlets are always emittance dominated, even in the presence of space charge, due to the charge suppression.
After a proper drift, a scintillator screen produces an image of the beamlets.
From the analysis of these beamlets it is possible to retrieve the transverse emittance in a single shot.
This system is widely used in the RF photoinjectors, at low energy, where the beam is space charge dominated.

However, several constraints must be considered in the design of the experiment.
Among others, the ratio of the size of the holes and the thickness of the material determines the angular acceptance of the device.
It must be large enough to avoid cutting the beam phase space.
The design of such a system, even at an energy range of several hundreds of MeV, much greater than the original design of the device, has been successfully demonstrated \cite{bib:pp}.

\begin{figure}[h]
  \centering
  \includegraphics[width=90mm]{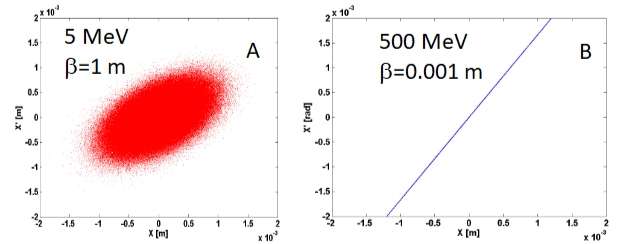}
  \caption{A: typical trace space from a RF gun; B: typical trace space from a plasma accelerator}
  \label{fig4}
  \end{figure}

The main problem related to such a diagnostic is in the sampling accuracy of the phase space.
The phase space of the beam emerging from a plasma acceleration is completely different with respect to the beam coming from a RF photoinjector, as pointed out in Ref.  \cite{bib:cianchi3}.
This main difference is shown in Fig. \ref{fig4}.

The phase space of a plasma accelerated beam is very thin and  strongly correlated.
This technique is a mapping of the phase space.
The sampling is mainly fixed by the geometry of the device.
When the phase space becomes very thin, the undersampling can yield completely wrong results.

\subsubsection{Betatron radiation}

Inside the plasma, the electron beam experiences several oscillations, called betatron oscillations.
Charged particles emit radiation when they are accelerated, and the radiation associated with these oscillations is called betatron radiation \cite{bib:betatron}.
In Ref. \cite{bib:plateu} there is an example of betatron spectroscopy. In the experiment, realized at the LOASIS facility of the Lawrence Berkeley National, a 400 MeV energy beam was used with a rms energy spread of less than 5$\%$ and 1 mrad divergence from a plasma density of 5 10$^{18}$ cm$^{-3}$.
The spectrum of the radiation was analysed in order to compare it with a simulated model, in order to reconstruct the beam emittance.
While with this method it was possible to measure emittance well below 1 mm-mrad, the technique is not able to measure the correlation term in the emittance formula.

In Ref. \cite{bib:kneip} the betatron radiation is used to produce Fresnel diffraction for a sharp edge.
A point-like beam produces a sharp profile, while a convolution between a step function and beam transverse distribution is expected otherwise. In this experiment, using also a dipole to bend the electron beam, several quantities are measured at the same time: the beam size, by means of betatron radiation; beam divergence, measuring the vertical dimension after a dipole; energy and energy spread with the same dipole. The correlation term is the only missed quantity needed to retrieve the value of the emittance, although
 the uncorrelated emittance can give in any case an upper limit for the total emittance.

\begin{figure}[h]
  \centering
  \includegraphics[width=90mm]{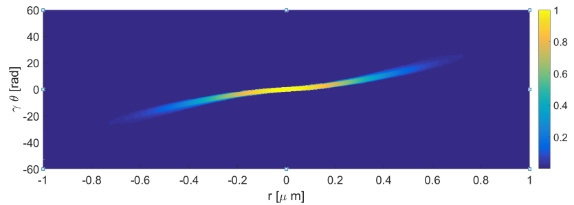}
  \caption{Reconstructed phase space with this technique. Laser parameters: energy 1 J, pulse duration 30 fs (FWHM), 10 $\mu$m diameter focus, a$_0$  $\approx$ 4.4; plasma density =(8$\pm$1)10$^{18 }$ cm$^{-3 }$.}
  \label{fig5}
  \end{figure}

Finally the emittance with the correlation term has been measured using  betatron radiation \cite{bib:curcio} and  Fig. \ref{fig5}  reports the measured phase space.

The main difference here is the simultaneous measurement of the beam energy spectrum and the betatron radiation energy spectrum.
Having in the same shot both of these quantities it is possible to reconstruct the phase space, in principle.
In the reported measurement the beam was self-injected, and so the initial parameters were not known.
For this reason some assumptions have to be made, resulting in an overestimation of the correlation term.
But if we can have a beam externally injected in the plasma, and it is possible to determine the full 6D phase space of such a beam and using it as starting point, no approximations are needed and the correlation term will be correctly evaluated. 
However, the betatron radiation has a severe drawback in the case of beam driven plasma accelerators.
The spectra coming from the driver beam and the witness bunch  overlap.
Moreover, the driver contains much more charge with respect to the witness and so even the tail of its spectrum can completely hide the witness spectrum.

\subsubsection{Single shot quadrupole scan}

\begin{figure}[h]
  \centering
  \includegraphics[width=90mm]{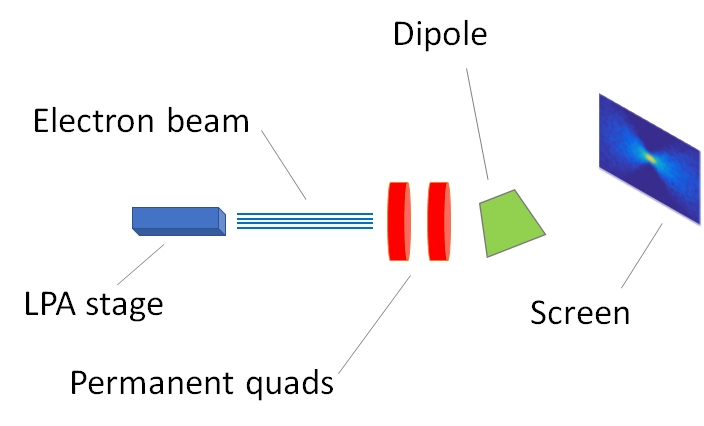}
  \caption{An electron beam produced by LPA stage is focused by a doublet of permanent magnets, before entering  a dipole. The dimension of the beam at different energy will change due to the energy dependence of the focusing power.}
  \label{fig6}
  \end{figure}

In the quadrupole scan technique the parameter that is under variation is the current in a quadrupole.
However, if we use permanent magnets, i.e., fixed magnetic field, and the beam has a remarkable energy spread, different parts of the beam will be focused at different spots.
Adding a dispersive element, like a dipole, will produce different beam sizes at different transverse positions.
A sketch of this system is shown in Fig. \ref{fig6}.

Starting from the first proposal \cite{bib:permquad1}, several groups have already implemented this technique  \cite{bib:permquad2,bib:permquad3}.
It seems very attractive for beam laser plasma acceleration (LPA).
However, there is a strong approximation in the data analysis.
There is the assumption that the emittance is the same at every energy.
This fact must be verified with simulations.
Also, this system needs a large energy spread (much larger than 1\%) to work properly.
It is suitable for a plasma accelerator in development and during the commissioning phase, while it is not feasible for a real machine, where the energy spread must be well below this value.
Other devices are still in the stage of early development, and the reader can find more details in Ref. \cite{bib:cianchi4}.

\subsection{Longitudinal measurement}

We have already talked about the TDS, even with a few fs resolution.
So, a diagnostic for very short beam and single shot is already available.
Why do we need to look for other solutions?
The main reasons are the cost of the device and the space needed.
X-band structures and X-band power sources are still very expensive.
A TDS, including the power source, the distribution waveguides, the cooling system can easily cost close to 1 M\euro.
 The length of such a device can be of the order of one metre or even longer.
So, this solution is not very compact and it is very expensive.

A bunch charge can emit coherent radiation at a wavelength longer than its length.
The characteristics of this radiation depends on the physical phenomena involved, such as synchrotron radiation, transition radiation, Smith--Purcell, and so on.
The analysis of the spectrum of such radiation can reveal the bunch longitudinal structure.
The spectrum intensity is given by

\begin{equation}
I_{\mathrm{tot}} \left( \omega  \right) = I_{\mathrm{sp}} \left( \omega  \right)\left[ {N + N\left( {N - 1} \right)F\left( \omega  \right)} \right]
\label{spectrum} \, ,
\end{equation}
where $I_{\mathrm{sp}}(\omega)$ is the spectrum produced by a single particle, $N$ is the number of particles, and $F$($\omega$) is the so-called form factor.
The physics of the emitting process is contained inside $I_{\mathrm{sp}}(\omega)$, while the information about the bunch is in $F(\omega)$.
The first term in the square bracket corresponds to incoherent emission and scales with the number of  particles, while the second term scales with $N^2$ and comes from the coherent emission.
The form factor can be expressed as a function of the charge distribution $\rho(z),$ as
\begin{equation}
F\left( \omega  \right) = \left| {\int\limits_{ - \infty }^\infty  {\rho \left( z \right)} \mathrm{e}^{\mathrm{i}\frac{\omega }{c}z} \mathrm{d}z} \right|^2 \, .
\label{formfactor}
\end{equation}

The inverse transformation gives the required value of $\rho \left( z \right)$
\begin{equation}
\rho \left( z \right) = \frac{1}{{\pi c}}\int\limits_0^\infty  {\sqrt {F\left( \omega  \right)} \cos \left( {\frac{\omega }{c}z} \right)}  \, .
\label{rho}
\end{equation}

With an even transformation, the odd terms in the longitudinal distribution are not considered and a phase reconstruction technique must be used to retrieve phase information \cite{bib:shibata}.
So the information about the longitudinal parameter is inside the form factor, which can be obtained from the power spectrum.
The principles of Fourier transform spectroscopy \cite{bib:fourier1, bib:fourier2} were discovered by Michelson and Rayleigh, who identified that the interference pattern from a two-beam interferometer, obtained by altering the path difference between the two beams, is the Fourier transform of the radiation passing through the interferometer,
\begin{equation}
I\left( \omega  \right) \propto \int\limits_{ - \infty }^\infty  {I\left( \delta  \right)} \cos \left( {\frac{{\omega \delta }}{c}} \right)\mathrm{d}\delta \, ,
\label{spectrum}
\end{equation}
where $\delta$ is the path length difference between the two interferometer arms: $I(\delta)$ is called the interferogram.

\begin{figure}[h]
  \centering
  \includegraphics[width=90mm]{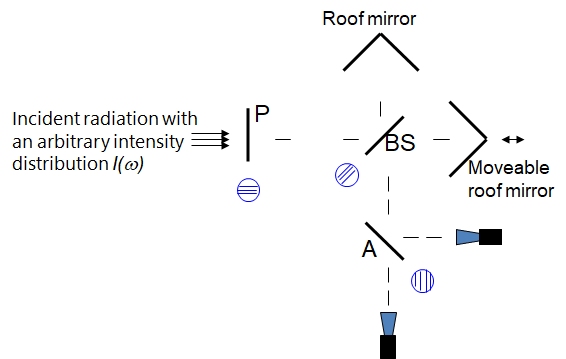}
  \caption{Sketch of a Martin--Puplett interferometer}
  \label{fig7}
  \end{figure}

An example of the device needed for such a measurement is shown in Fig. \ref{fig7}.

This interferometer is of Martin--Puplett type \cite{bib:MP}, the polarizing version of the Michelson one, where the Michelson beam splitter is replaced by a polarizer which splits the polarizations and plane mirrors by roof mirrors.
The radiation, coming from a plane mirror, enters the first polarizer (P) whose wires, in this case, are horizontal to let vertical polarization through.
The vertically polarized transmitted radiation first reaches the beam splitter (BS), placed at 45 degrees, whose wires are at 45 degrees  to the horizontal plane when viewed along the beam input axis.
The BS splits the input signal into two equal, orthogonal polarized components, one being reflected towards the stationary roof mirror and the other being transmitted to the movable one.
Radiation coming back from both arms to the beam splitter recombines and is focused onto the detectors by a second parabolic mirror via the analysing grid (A), located between the mirror and the detectors, which transmits the component orthogonal to the wires to one detector and reflects the component parallel to the wires to the other.

So to estimate the bunch length we need to measure the spectrum of the coherent radiation, and to do this we have to record the intensity of the radiation on a detector in an interferometer, varying the path difference between two arms of this device.
Since the bunch lengths for high brightness beams are of the order of picoseconds or even shorter, the coherent radiation is usually in the terahertz frequency range.
For such frequencies, the most used detectors are thermal detectors, like Golay cells \cite{bib:golay} and pyroelectric detectors.

One of the main issues in this kind of measurement is the precise reconstruction of the entire spectrum,  as there is a low frequency cutoff owing to the vacuum pipe, the interferometer components, and the detector acceptance.
The overall transfer function of the whole system must be known, including the transfer function of the vacuum window and the transport line.
For all these reasons, a practical approach to this measurement is to directly fit the spectrum with well-known guessed distributions \cite{bib:murokh}.

Longitudinal diagnostics with this technique began about 20 years ago and they are now widespread.
In terms of compactness it is very attractive because only a small metallic target is needed in the beam line.
Also, it is a very cheap diagnostic, compared with TDS structure.
But it has the big disadvantage of being a multi-shot method.

However, some groups have already performed single shot measurements.
In \Bref{bib:wesch}, Wesch \textit{et al.} used a multi-stage spectrometer with a series of blazed reflection gratings.
At any stage, wavelengths shorter than a threshold are dispersed while the longer wavelengths are reflected.
With every grating focused on an array of detectors, it is possible to acquire in a single shot the whole spectrum for terahertz and infrared spectroscopy.
The main challenge of such a device is the alignment of the various stages.
An improved version of this device, using separate spectrometers, each one looking at different wavelengths, has been successfully tested on a plasma accelerated beam \cite{bib:matthias}.

In Ref. \cite{bib:maxwell} Maxwell \textit{et al.} used a single-stage single-shot device, based on a KRS-5 (thallium bromoiodide) prism.
This material is quite interesting, even if it is hygroscopic, because it has a quite flat response between 0.8 and $40 \Uum$.
The radiation is dispersed and then sent to 128 lead zirconate titanate pyroelectric elements with a $100 \Uum$ spacing line array.

The major advantage of these techniques based on  coherent radiation is the time resolution. The emitted wavelength scales as the bunch length. So the only requirement is a device sensitive to such wavelengths.
For instance, $1 \Ups$ means $300 \Uum$, while $1 \Ufs$ needs a wavelength sensitivity of $300 \Unm$, where a lot of detectors are available.

However, like all the techniques based on the reconstruction of the spectrum, not linear transmission at all the frequencies, absorption, frequency filtering due to geometrical constraints, and all the other effects that can modify the spectrum can affect noticeably the measurement.

\section{Conclusions}

The development of plasma accelerators  not only focuses on accelerating structures. Plasma lenses \cite{bib:pompili}, a plasma dechirper \cite{bib:vladimir}--\cite{bib:darcy} are already real devices, while there are ideas about plasma deflectors \cite{bib:pladef}, plasma dipoles, and plasma dumps.
The diagnostics of conventional machines must be redesigned completely, in order to reduce their dimensions.
At the same time new solutions must be found to measure with single shot capability both longitudinal and transverse parameters.
Plasma accelerators are the right place for all  students who are discouraged because they believe that everything has already been done and that progress can be only marginal, to demonstrate that there is a long way to go, before moving to a museum the present particle accelerators.

\section{Acknowledgments}

I would like to thank Enrica Chiadroni and Andrea Mostacci for  useful discussions and  help in revising the document.

\end{document}